\pdfoutput=1
\documentclass{tlp}

\usepackage[utf8]{inputenc}
\usepackage[T1]{fontenc}
\usepackage[english]{babel}%

\usepackage{hyperref}
\usepackage{qsymbols,xparse}
\usepackage{amsfonts,amsmath}
\usepackage{nicefrac}
\usepackage{float}

\renewcommand\footnotemark{}

\newcommand{\cgL}{\mathcal{L}}
\newcommand{\D}{\mathcal{D}}

\usepackage{graphicx}
\usepackage{url}
\usepackage{verbatim}
\usepackage{color}
\definecolor{lgray}{gray}{0.95}
\definecolor{lblue}{rgb}{0.90,0.90,1.00}
\definecolor{lyellow}{rgb}{1.00,1.00,0.70}

\usepackage{listings}
\newtheorem{ex}{Example}

\newenvironment{codex}{\small\verbatim}{\endverbatim\normalsize}

\lstnewenvironment{code}
    {\lstset{}%
      \csname lst@SetFirstLabel\endcsname}
    {\csname lst@SaveFirstLabel\endcsname}
\newcommand{\BI}[0]{\begin{itemize}}
\newcommand{\EI}[0]{\end{itemize}}

\newcommand{\BE}[0]{\begin{enumerate}}
\newcommand{\EE}[0]{\end{enumerate}}

\newcommand{\BX}[0]{\begin{ex}}
\newcommand{\EX}[0]{\end{ex}}
\newcommand{\BV}[0]{\begin{verbatim}}
\newcommand{\EV}[0]{\end{verbatim}}

\def \bscale1 {0.25}
\def \bscale {0.25}

\newcommand{\FIG}[4]{
\begin{figure}[htbp!]
\centering
{\includegraphics[scale=#3]{./#4}}
\caption{#2}
\label{#1}
\end{figure}
}

\begin{document}

\title{Random generation of closed simply-typed $`l$-terms:
a synergy between logic programming and Boltzmann samplers}

\author[Maciej Bendkowski, Katarzyna Grygiel and Paul Tarau]
{Maciej Bendkowski, Katarzyna Grygiel\\
Theoretical Computer Science Department\\
  Faculty of Mathematics and Computer Science\\
  Jagiellonian University\\
  ul. Prof. {\L}ojasiewicza 6, 30-348 Krak\'ow, Poland\\
  \email{\{bendkowski,grygiel\}@tcs.uj.edu.pl}
  \and Paul Tarau\\
  Department of Computer Science and Engineering\\
University of North Texas\\
Denton, TX, USA\\
  \email{paul.tarau@unt.edu} \thanks{The first two authors have been partially supported by the Polish National Science Center grant 2013/11/B/ST6/00975. The third author has been supported by NSF grant 1423324.
  A conference version of the present paper appears in the proceedings
  of PADL 2017 (PC chairs: Yuliya Lierler and Walid Taha).}}

\maketitle

\begin{abstract}
A natural approach to software quality assurance consists in writing unit tests securing programmer-declared code invariants. Throughout the literature a great body of work has been devoted to tools and techniques automating this labour-intensive process. A prominent example is the
successful use of randomness, in particular random typable $`l$-terms, in testing functional programming compilers such as the Glasgow Haskell Compiler.
Unfortunately, due to the intrinsically difficult combinatorial structure of typable $`l$-terms no effective  uniform sampling method is known, setting it as a fundamental open problem in the random software testing approach.
In this paper we combine the framework of Boltzmann samplers, a powerful technique of random combinatorial structure generation, with today's Prolog systems offering a synergy between logic variables, unification with occurs check and efficient backtracking. This allows us to develop a novel sampling mechanism able to construct uniformly random closed simply-typed $`l$-terms of up size 120.
We apply our techniques to the generation of uniformly random closed simply-typed normal forms and design a
parallel execution mechanism pushing forward the achievable term size to 140.
\end{abstract}

\keywords{Boltzmann samplers,
random generation of simply-typed $`l$-terms,
type inference,
combinatorics of $`l$-terms,
random generation of simply-typed normal forms,
parallel implementation of Boltzmann samplers}.

\section{Introduction}
Simply-typed $`l$-terms~\cite{hindley2008lambda,bar93} constitute the theoretical foundations of modern functional programming languages, such as Haskell or OCaml.
Types in $`l$-calculus provide an additional safety layer as typable $`l$-terms are necessarily strongly normalizing, i.e.~terminate in all evaluation orders, hindering the programmer from introducing some easy to avoid software bugs. Moreover, via the famous Curry-Howard isomorphism, closed $`l$-terms that are inhabitants of simple types
can be seen as proofs for tautologies in the implicational
fragment of minimal logic.

In~\cite{palka11} the authors used random typable $`l$-terms as a tool for testing the prominent Glasgow Haskell Compiler (GHC). Though successful for the purpose of finding optimisation bugs in GHC, their random terms were not \emph{uniformly} random with respect to size. In other words, some kinds of typable $`l$-terms were favoured over other kinds of equal size terms. Uniform generation, on the other hand, assigns equal probability to terms of equal size and hence produces `typical' typable $`l$-terms, without introducing an unintended nor explicit bias in the sampling process.

%Simply-typed $`l$-terms~\cite{hindley2008lambda,bar93} enjoy
%a number of nice properties, such as strong normalization,
%i.e.~termination for all evaluation-orders,
%a Cartesian closed category mapping and a set-theoretical
%semantics. More importantly, via the Curry-Howard isomorphism,
%closed $`l$-terms that are {\em inhabitants} of simple types
%can be seen as proofs for tautologies in the implicational
%fragment of {\em minimal logic} which, in turn, correspond to
%the simple types.
%Extended with a fix-point operator, simply-typed $`l$-terms
%can be used as the intermediate language for compiling
%Turing-complete functional languages.

Recent work on the combinatorics of
$`l$-terms~\cite{grygielGen,bodini11,normalizing13},
relying on generating functions and techniques from analytic
 combinatorics~\cite{flajolet09}, has provided counts for several families
of $`l$-terms and clarified important
quantitative properties of interesting subclasses of
$`l$-terms. With the techniques provided
by generating functions~\cite{flajolet09},
it was possible to separate the
{\em counting} of the terms of a given size for
several families of $`l$-terms from
their more computation intensive  {\em generation}, resulting
in several additions (e.g., {\bf A220894}, {\bf A224345}, {\bf A114851}) to
the On-Line Encyclopedia of Integer Sequences~\cite{intseq}.

On the other hand, due to the intricate interaction
between type inference and the applicative structure
of $`l$-terms, the combinatorics
of simply-typed $`l$-terms has left important
problems open. For instance, the basic problem of
counting the number of closed simply-typed
$`l$-terms of a given size.
At this point, obtaining counts for simply-typed
$`l$-terms requires going through the more computation-intensive
generation process.

Fortunately, by taking advantage of the synergy between logic variables,
unification with occurs check and efficient backtracking it
is possible to significantly accelerate the
generation of simply-typed lambda
terms~\cite{padl15} by interleaving it with type inference
steps. While the generators described in the aforementioned paper can push the size of the
simply-typed $`l$-terms by a few steps higher,
one may want to obtain uniformly sampled random terms of
significantly larger size,
especially if one is concerned
not only about correctness but also about
scalability of compilers and program transformation tools
used in the implementation of functional programming
languages  and proof assistants.

This brings us to the main contribution of this paper.
We will first build efficient generators for simply-typed
$`l$-terms that work by interleaving term building and
type inference steps.
From them, we will derive
Boltzmann samplers returning
random simply-typed $`l$-terms~\cite{grygiel2015} of sizes
between 120 and 140, assuming a slight variation of the
`natural size' introduced in~\cite{maciej16}, assigning to
each constructor a size given by its arity.
We will also extend this technique to the random generation
of simply-typed closed normal forms, based on the same definition
of size.

The paper is organized as follows.
Section~\ref{lgen} describes generators for plain, closed and simply-typed terms of a given size.
Section~\ref{anal-comb} revisits key notions from analytic combinatorics and the general design of Boltzmann samplers.
Section~\ref{bol} derives Boltzmann samplers for random generation of simply-typed closed $`l$-terms.
Section~\ref{nfgen} describes generators for $`l$-terms in normal form as well as their closed and simply-typed subsets.
Section~\ref{nfbol} derives Boltzmann samplers for random generation of simply-typed closed $`l$-terms in normal form.
Section~\ref{par} describes a simple parallel execution model using multiple independent threads.
Section~\ref{disc} discusses techniques for possibly pushing higher the sizes of generated random terms.
Finally, section~\ref{rels} overviews related work and section~\ref{concl} concludes the paper.

The paper is structured as a literate Prolog program.
The code has been tested with SWI-Prolog 7.3.8 and YAP 6.3.4.
It is also available as a separate file at
\url{http://www.cse.unt.edu/~tarau/research/2016/bol.pro}.

A limited conference version of the current paper appeared as~\cite{padl2017}.
We remark that in its following extended version, we include a full discussion
regarding the framework of Boltzmann samplers and its particular application to
random generation of closed simply-typed $`l$-terms, as well as a novel parallel
execution model supported with experimental results on a 44-core machine.

\begin{codeh}
% TPLP
%
:-ensure_loaded(library(lists)).
\end{codeh}

\section{Generators for $`l$-terms of a given natural size} \label{lgen}

We start by generating all $`l$-terms of a given size, in the de~Bruijn notation.

\subsection{De~Bruijn notation}
%\footnote{A $`l$-term is called {\em closed} if it has no free variables and {\em open} otherwise. A term is called {\em plain} if it is either closed or open.}

De~Bruijn indices~\cite{dbruijn72} provide a robust {\em name-free} representation of lambda
term variables. Closed terms, i.e.~terms without free variables,
that are identical up to renaming of variables share a unique representation. This allows each variable occurrence to be replaced by a non-negative integer marking the number of lambda abstractions between the
 variable and its binder.
 Following~\cite{maciej16} we assume a unary notation of integers using the constant {\tt 0} and the constructor {\tt s/1} for the successor. Lambda
  abstraction and application constructors are represented using {\tt l/1} and {\tt a/2},
  respectively. And so, the set $\cgL$ of {\em plain $`l$-terms} is given by the following grammar:
  \[ \cgL  = \cgL\, \cgL \ \mid\ `l\, \cgL \ \mid\ \D, \]
where $\D$ denotes the set $\{{\verb+0+}, {\verb+s(0)+}, {\verb+s(s(0))+},\ldots\}$ of de~Bruijn indices.

Throughout the paper we assume that each constructor is of \emph{weight} equal to its arity and the \emph{size} of a $`l$-term is the sum of the weights of its building constructors. Moreover, for simplicity we refer to set of unconstrained $`l$-terms, i.e.~either closed or open ones, as {\em plain} $`l$-terms.

\subsection{Generating plain $`l$-terms}

Generation of plain $`l$-terms of a given size proceeds by consuming
at each step a size unit, represented by the constructor {\tt s/1}.
This ensures that, for a size definition allocating a number of size units to each
of the constructors of a term, generation is constrained to terms of a given size.
As there are $n+1$  leaves (labeled {\tt 0}) in a tree with $n$ {\tt a/2} constructors,
we implement our generator to consume as many size-units as
the arity of each constructor, in particular $0$ for {\tt 0} and $2$
for the constructor {\tt a/2}. This means that we will obtain the counts
for terms of natural size $n+1$ when consuming $n$ size-units.\\

\noindent {\em {\tt genLambda(S,X)} is true if {\tt X} is is a plain lambda term of natural size {\tt S} where {\tt S} is a natural number in successor arithmetic.}

\begin{code}
genLambda(s(S),X):-genLambda(X,S,0).

genLambda(X,N1,N2):-nth_elem(X,N1,N2).
genLambda(l(A),s(N1),N2):-genLambda(A,N1,N2).
genLambda(a(A,B),s(s(N1)),N3):-
  genLambda(A,N1,N2),
  genLambda(B,N2,N3).
\end{code}
Note that {\tt nth\_elem/3} consumes progressively larger size-units for variables
of a higher de~Bruijn index, a property that conveniently mimics the fact that, in practical
programs,  variables located farther from their binders are likely to occur
less frequently than those
closer to their binders.
\begin{code}
nth_elem(0,N,N).
nth_elem(s(X),s(N1),N2):-nth_elem(X,N1,N2).
\end{code}

\BX
Plain $`l$-terms of size 2 (where the size of each constructor is given by its arity).
\begin{codex}
?- genLambda(s(s(s(0))),X).
X = s(s(0)) ; X = l(s(0)) ; X = l(l(0)) ; X = a(0, 0) .
\end{codex}
\EX
Counts for plain $`l$-terms are given by the sequence {\bf A105633} in~\cite{intseq}.

\subsection{Generating closed $`l$-terms} \label{lambda}

We derive a generator for closed $`l$-terms by
counting with help of a list of logic variables.
At each lambda binder {\tt l/1} step, a new variable is added to the list
associated with a path from the root. For now, we simply use
the {\em length of the list} as a counter for {\tt l/1} nodes
on the path. By ensuring that de Bruijn indices are less or
equal to the number of lambdas on the path from a leaf to the root
we ensure that only closed terms are generated.
Later, we will use the actual variables to store the type
of the corresponding lambda binders when implementing type inference.

The predicate {\tt genClosed/2}
builds this list of logic variables as it generates binders.
When generating a leaf variable,
it picks `non-deterministically' one of the variables  among
the list of variables corresponding to binders encountered on a given path from the root
{\tt Vs}. In fact, this list of variables will be ready to be used later to store
the types inferred for a given binder.\\

\noindent {\em {\tt genClosed(S,X)} is true if {\tt X} is is a closed lambda term of natural size {\tt S} where {\tt S} is a natural number in successor arithmetic.}

\begin{code}
genClosed(s(S),X):-genClosed(X,[],S,0).

genClosed(X,Vs,N1,N2):-nth_elem_on(X,Vs,N1,N2).
genClosed(l(A),Vs,s(N1),N2):-genClosed(A,[_|Vs],N1,N2).
genClosed(a(A,B),Vs,s(s(N1)),N3):-
  genClosed(A,Vs,N1,N2),
  genClosed(B,Vs,N2,N3).
\end{code}
Like {\tt nth\_elem} in the case of plain $`l$-terms,
 the predicate {\tt nth\_elem\_on} consumes progressively
larger size-units for variables of a higher de~Bruijn index.
At the same time, its second (list) argument
ensures that on each branch leading from a leaf
to the root of the tree, there is a variable
introduced by a lambda binder above it. This gets our code ready
for the next refinement, where we will use these variables to
store our inferred types.
\begin{code}
nth_elem_on(0,[_|_],N,N).
nth_elem_on(s(X),[_|Vs],s(N1),N2):-nth_elem_on(X,Vs,N1,N2).
\end{code}

\BX
Closed $`l$-terms of natural size 5.
\begin{codex}
?- genClosed(s(s(s(s(s(0))))),X).
X = l(l(l(s(0)))) ; X = l(l(l(l(0)))) ; X = l(l(a(0, 0))) ;
X = l(a(0, l(0))) ; X = l(a(l(0), 0)) ; X = a(l(0), l(0)) .
\end{codex}
\EX
Counts for closed $`l$-terms are given by the sequence {\bf A275057} in~\cite{intseq}.

\subsection{Generating simply-typed $`l$-terms}

We will derive a generator for simply-typed $`l$-terms from the generator
for closed terms. The list of variables added there to ensure that
each index has binder will be used to contain the types, on which de~Bruijn indices
pointing to the same binder should agree. Note that we use the right associative
infix constructor ``\verb~->~'' to denote the arrows connecting simple types.\\

\noindent {\em
{\tt genTypable(X,V,Vs,N1,N2)} holds if {\tt X} is well-typed with type {\tt V}
in type environment {\tt Vs} and has size {\tt N1-N2} computed in
successor arithmetic notation
}

\begin{code}
genTypable(X,V,Vs,N1,N2):-genIndex(X,V,Vs,N1,N2).
genTypable(l(A),(X->Xs),Vs,s(N1),N2):-genTypable(A,Xs,[X|Vs],N1,N2).
genTypable(a(A,B),Xs,Vs,s(s(N1)),N3):-
  genTypable(A,(X->Xs),Vs,N1,N2),
  genTypable(B,X,Vs,N2,N3).
\end{code}
The predicate {\tt genIndex/5} ensures, via {\em unification with occurs-check},
that the same non-cyclic type is assigned to each leaf corresponding to an occurrence
of  a variable introduced by a given lambda binder.\\

\noindent {\em
{\tt genIndex(K,V,Vs,N1,N2)} holds if at position {\tt K} type {\tt V} is found
in type environment {\tt Vs} and the available size resource has been reduced as {\tt N2=N1-K}, computed in
successor arithmetic notation
}
\begin{code}
genIndex(0,V0,[V|_],N,N):-unify_with_occurs_check(V0,V).
genIndex(s(X),V,[_|Vs],s(N1),N2):-genIndex(X,V,Vs,N1,N2).
\end{code}

We expose this algorithm via two interfaces: one for
plain terms and one for closed terms. Their only difference is
the constraint that the list of available variables for closed
terms is initially empty.\\

\noindent {\em {\tt genPlainTypable(S,X,T)} is true if {\tt X} is is a plain simply-typed lambda term of natural size {\tt S} where {\tt S} is a natural number in successor arithmetic and {\tt T} is the type inferred for {\tt X}.}\\

\noindent {\em {\tt genClosedTypable(S,X,T)} is true if {\tt X} is is a  closed simply-typed lambda term of natural size {\tt S} where {\tt S} is a natural number in successor arithmetic and {\tt T} is the type inferred for {\tt X}.}\\

\begin{code}
genPlainTypable(S,X,T):-genTypable(S,_,X,T).

genClosedTypable(S,X,T):-genTypable(S,[],X,T).

genTypable(s(S),Vs,X,T):-genTypable(X,T,Vs,S,0).
\end{code}
For convenience, we shift the sequence by one to match the
size definition where both application nodes and
{\tt 0} leaves have size $1$ as originally given
in \cite{maciej16}. As there are $n+1$ leaf nodes for $n$ application
nodes, consuming two units for an application rather than one
for an application and one for a leaf as done in \cite{maciej16},
speeds up the generation process as we are able to apply the
size constraints at application nodes,
earlier in the recursive descent.

\BX
Plain simply-typed $`l$-terms of natural size 3.
\begin{codex}
?- genPlainTypable(s(s(s(s(0)))),X,T).
X = s(s(s(0))),T = A ;
X = l(s(s(0))),T =  (A->B) ;
X = l(l(s(0))),T =  (A->B->A) ;
X = l(l(l(0))),T =  (A->B->C->C) ;
X = a(0, s(0)),T = A ;
X = a(0, l(0)),T = A ;
X = a(s(0), 0),T = A ;
X = a(l(0), 0),T = A .
\end{codex}
\EX
Counts for plain simply-typed $`l$-terms, up to size 16, are given by the sequence:
\[
0,1,2,3,8,17,42,106,287,747,2069,5732,16012,45283,129232,370761,1069972.
\]
Counts for closed simply-typed $`l$-terms are given
by the sequence {\bf A272794} in
\cite{intseq}. The first 17 entries are:
\[
0,0,1,1,2,5,13,27,74,198,508,1371,3809,10477,29116,82419,233748.
\]

\section{Analytic combinatorics}\label{anal-comb}
Our approach to random $`l$-term generation relies on the powerful theory of \emph{analytic combinatorics} and, in particular, the design of Boltzmann samplers.
In this section we excerpt the main ideas and notions used throughout the remainder of the paper. We refer the curious reader to~\cite{flajolet09,Wilf:2006:GEN:1204575} for a detailed exposition on generating functions, the singularity analysis process and its applications, as well as~\cite{Duchon:2004} for a reference on Boltzmann samplers in general.

\subsection{Symbolic method}

Let $\mathcal{A}$ be a denumerable set of combinatorial objects, e.g.~inhabitants of an algebraic data type. Suppose that $\mathcal{A}$ is additionally equipped with a \emph{size notion} $|\cdot| \colon \mathcal{A} \to \mathbb{N}$, assigning each $\alpha \in \mathcal{A}$ its size $|\alpha|$ in such a way that for each $n \in \mathbb{N}$ there are only finitely many objects in $\mathcal{A}$ of size $n$. Then, $\mathcal{A}$ together with $|\cdot|$ form a \emph{combinatorial class} -- the central object in the theory of \emph{analytic combinatorics}~\cite{flajolet09}. Through analytic combinatorics we obtain systematic methods of creating, manipulating and studying the behaviour of combinatorial structures, in particular, the properties of large typical objects as well as their effective random generation.

Let $a_n$ denote the number of objects in $\mathcal{A}$ of size $n$. Then, ${(a_n)}_{n \in \mathbb{N}}$ becomes the \emph{counting sequence} of $\mathcal{A}$. Suppose that we assign a formal power series $A(z)$ to ${\mathcal A}$'s counting sequence in such a way that $a_n$ becomes $A(z)$'s $n$th coefficient (denoted $[z^n]A(z)$), i.e.
\[ A(z) = \sum_{n \geq 0} a_n z^n. \]
The series $A(z)$ is called then the \emph{generating function} of $\mathcal{A}$ (see, e.g.~\cite{Wilf:2006:GEN:1204575}).

Such a compact representation of $\mathcal{A}$'s counting sequence enjoys a number of elegant manipulation properties. Suppose we have two disjoint combinatorial classes $\mathcal{A}$ and $\mathcal{B}$ and wish to construct a combinatorial class $\mathcal{C} = \mathcal{A} + \mathcal{B}$ consisting of the union of $\mathcal{A}$ and $\mathcal{B}$ with unmodified size functions. Then, the generating function $C(z)$ of $\mathcal{C}$ is given by
\[ C(z) = \sum_{n \geq 0} (a_n + b_n) z^n. \]
Hence, using the formal series addition,
\[ C(z) = A(z) + B(z). \]

Now, suppose that we wish to construct a combinatorial class $\mathcal{C} = \mathcal{A} \times \mathcal{B}$ consisting of \emph{pairs} in the form of $(\alpha,\beta)$ where $\alpha \in \mathcal{A}$, $\beta \in \mathcal{B}$ and the size of $(\alpha,\beta)$ is the sum of $\alpha$'s and $\beta$'s sizes. In other words, on the level of generating functions
\[ C(z) = \sum_{n \geq 0} \sum_{k=0}^{n} (a_{n-k} b_k) z^n. \]
Note that this is precisely the Cauchy product formula, therefore
\[ C(z) = A(z) B(z). \]

The above, so called \emph{admissible constructions}, allow us to find generating functions for a broad class of algebraic data types, including $`l$-terms in the de~Bruijn notation, by means of the following \emph{symbolic method}~\cite{flajolet09}.

Let us start with de~Bruijn indices. Recall that de~Bruijn indices are defined by the following grammar
\[ \mathcal{D} = 0~|~S(\mathcal{D}). \]
Note that on the right-hand side we have two disjoint sets of indices -- a singleton class consisting of $0$ and a class of successors. Since zero is of size $0$ (as it is a constant), its corresponding generating function is simply equal to $z^0 = 1$. The class of successors, on the other hand, is a bit more involved. Suppose that we construct an auxiliary class $\mathcal{S}$ consisting of a single object \verb+s+ of size $1$. Then, we can bijectively assign to each successor in $\mathcal{D}$ a pair from $\mathcal{S} \times \mathcal{D}$. In consequence, we obtain the following functional equation on $\mathcal{D}$'s generating function $D(z)$
\begin{eqnarray*}
 D(z) &=& 1 + z D(z)\\
 &=& \frac{1}{1-z}.
\end{eqnarray*}

Now, let us consider the class $\mathcal{L}$ of $`l$-terms, given by the following grammar
\[ \mathcal{L} = \mathcal{D}~|~\lambda \mathcal{L}~|~\mathcal{L} \mathcal{L}. \]
Again, on the right-hand side we have three disjoint combinatorial classes -- the set of de~Bruijn indices, the set of $`l$-terms starting with an abstraction, and finally the class of term applications. Using the same symbolic method as for $\mathcal{D}$, we obtain the following functional equation defining $L(z)$ (recall that the abstraction is of size $1$ whereas the application is of size $2$):
\begin{eqnarray*}
L(z) &=& D(z) + z L(z) + z^2 {L(z)}^2\\
 &=& \frac{1}{1-z} + z L(z) + z^2 {L(z)}^2.\\
\end{eqnarray*}
Solving the above quadratic equation in $L(z)$ we obtain two possible solutions:
\[ L(z) = \frac{1-z+\frac{\sqrt{1-3z-z^2-z^3}}{\sqrt{1-z}}}{2 z^2} \quad \text{ or } \quad
   L(z) = \frac{1-z-\frac{\sqrt{1-3z-z^2-z^3}}{\sqrt{1-z}}}{2 z^2}. \]
Note that since there exists just a single $`l$-term of size equal to $0$ (the de~Bruijn index \verb+0+), we expect that $\lim_{z \to 0} L(z) = [z^0]L(z) = 1$. This condition holds only for the latter equation, hence finally
\begin{equation}\label{eq:L(z)-def}
L(z) = \frac{1-z-\frac{\sqrt{1-3z-z^2-z^3}}{\sqrt{1-z}}}{2 z^2}.
\end{equation}

\subsection{Singularity analysis}

Although generating functions, as discussed previously, are formal series, \emph{analytic combinatorics}~\cite{flajolet09} links their analytic properties, when viewed as complex functions in one variable $z$, with the properties of the underlying counting sequences.
Surprisingly profound questions regarding the asymptotic behaviour and statistical properties of the underlying counting sequences might be addressed by carefully examining
the \emph{dominant singularities} of the corresponding generating functions (so called \emph{singularity analysis}).

And so, for a broad class of combinatorial classes, including $`l$-terms in the de~Bruijn notation, it is possible to give accurate approximations on the number of objects of size $n$, or investigate the properties of large random structures. In~\cite{maciej16} the authors gave the following asymptotic approximation of $[z^{n+1}]L(z)$ (note that $[z^{n+1}]L(z)$ in their size notion is equal to $[z^n]L(z)$ as given by Equation~\eqref{eq:L(z)-def}):

\[ [z^{n+1}]L(z) \sim {\left(\frac{1}{\rho}\right)}^n \frac{C}{n^{3/2}}, \]
where $\rho \approx 0.29560$ (hence $\nicefrac{1}{\rho} \approx 3.38298$) and $C \approx 0.60676$. Here, $\rho$ is the \emph{dominant singularity} of $L(z)$, i.e.~the radius of convergence of $L(z)$. %when viewed as a complex function.

Note that in the case of $L(z)$, the location of $\rho$ dictates the exponential rate of growth of $[z^{n+1}]L(z)$. The precise nature and neighbourhood of $\rho$ determine the sub-exponential factor $\frac{C}{n^{3/2}}$. For the purpose of this paper, we are interested in the approximation of $\rho$ and the evaluation of $L(z)$ in arbitrary parameters from the interval $(0, \rho)$, as we use them in the construction of a rejection Boltzmann sampler for $`l$-terms.

\subsection{Boltzmann samplers}
In their breakthrough paper~\cite{Duchon:2004}, Duchon et al. introduced a powerful framework of \emph{Boltzmann samplers} meant for random generation of combinatorial structures. Suppose we have a generating function
\[ A(z) = \sum_{n \geq 0} a_n z^n. \]
We wish to design an efficient algorithm, which returns a random structure $\alpha \in \mathcal{A}$ in such a way that any two structures of equal size have the same probability of being chosen. In other words, we want the probability $\mathbb{P}(\alpha)$ that $\alpha \in \mathcal{A}$ of size $n$ is the sampler's outcome to be equal to
\[ \mathbb{P}(\alpha) = \frac{1}{a_n}. \]

Duchon et al. proposed the following approach. Suppose we relax our restriction that the sampler's outcome size is deterministic and parametrize the sampler with an additional real parameter $x \in (0, \rho)$ where $\rho$ is the dominating singularity of $A(z)$.
Let us set a probability space on $\mathcal{A}$ such that $\mathbb{P}_{x}(\alpha)$, the probability that $\alpha \in A$ is the sampler's outcome, is equal to
\[ \mathbb{P}_{x}(\alpha) = \frac{x^{|\alpha|}}{A(x)}. \]

 Let $N$ be the random variable marking the size of the sampler's outcome. Then, the probability $\mathbb{P}_{x}(N = n)$ that the sampler returns an object of size $n$ is equal to
\[ \mathbb{P}_{x}(N = n) = \frac{a_n x^n}{A(x)}. \]
Note that this is indeed a probability since
\[ \sum_{n \geq 0} \mathbb{P}_{x}(N = n) = \frac{1}{A(x)} \sum_{n \geq 0} a_n x^n = 1. \]
We can therefore consider the \emph{expected} outcome size and all its higher moments. In particular, it is easy to verify that the expected size $\mathbb{E}_x(N)$ and the standard deviation $\sigma_x(N)$ are given by
\begin{equation}\label{eq:E(N)}
\mathbb{E}_x(N) = x \frac{A'(x)}{A(x)} \quad \text{and} \quad \sigma_x(N) = \sqrt{ \frac{x^2 A''(x) + x A'(x)}{A(x)} - {\left(x \frac{A'(x)}{A(x)}\right)}^2 }.
\end{equation}

Hence, in this model we do not control the exact size of the sample, although we can \emph{calibrate} its expected size and standard deviation by choosing a suitable parameter $x$.

\subsection{Constructing Boltzmann samplers}
Let $\mathcal{A}$ be a combinatorial class for which we want to design a Boltzmann sampler $\Gamma_x(\mathcal{A})$. The process of constructing $\Gamma_x(\mathcal{A})$ described by Duchon et al.~\cite{Duchon:2004} follows the recursive structure of $\mathcal{A}$.

Suppose that $\mathcal{A} = \mathcal{B} + \mathcal{C}$. Let $\alpha \in \mathcal{A}$. Since both $\mathcal{B}$ and $\mathcal{C}$ are disjoint, the probabilities $\mathbb{P}_{\Gamma, x}(\alpha \in \mathcal{B})$ that $\alpha \in \mathcal{B}$ and $\mathbb{P}_{\Gamma, x}(\alpha \in \mathcal{C})$ that $\alpha \in \mathcal{C}$ are equal to
\[ \mathbb{P}_{\Gamma, x}(\alpha \in \mathcal{B}) = \frac{B(x)}{A(x)} \quad \text{and} \quad \mathbb{P}_{\Gamma, x}(\alpha \in \mathcal{C}) = \frac{C(x)}{A(x)}. \]

It means therefore that in order to sample an object from $\mathcal{A}$ we have to make a probabilistic decision which branch, i.e.~$\mathcal{B}$ or $\mathcal{C}$, to choose. We draw uniformly at random a real $r \in [0,1]$ and compare it with the \emph{branching probabilities} $\frac{B(x)}{A(x)}$ and $\frac{C(x)}{A(x)}$. Then, we call recursively one of the corresponding samplers $\Gamma_x(\mathcal{B})$ or $\Gamma_x(\mathcal{C})$, continuing the sampling process.

Now, suppose that $\mathcal{A} = \mathcal{B} \times \mathcal{C}$. Let $\alpha = (\beta,\gamma) \in \mathcal{A}$. Note that
\[\mathbb{P}_{\Gamma, x}(\alpha \in \mathcal{A}) = \frac{x^{|\alpha|}}{A(x)}
= \frac{x^{|\beta| + |\gamma|}}{B(x) C(x)} = \mathbb{P}_{\Gamma, x}(\beta \in \mathcal{B})  \cdot \mathbb{P}_{\Gamma, x}(\gamma \in \mathcal{C}).\]
In other words, in order to sample an object from $\mathcal{A}$ we have to sample two objects -- independently one from $\mathcal{B}$ and one from $\mathcal{C}$ -- and make a pair out of them.

The recursion stops at the level of singleton classes. In such a case, we simply return the single object in our class, since
\[ \mathbb{P}_{\Gamma, x}(\alpha \in \mathcal{A}) = \mathbb{P}_x(\alpha) = \frac{x^{|\alpha|}}{A(x)} = \frac{x^{|\alpha|}}{x^{|\alpha|}} = 1. \]

We note that tough the framework of Boltzmann samplers is much more involved,
the above two design patterns (essentially instances of so-called datatype
generic programming) allow us to easily construct a Boltzmann sampler for plain
$`l$-terms, exploiting the generating function $L(z)$, see
Equation~\eqref{eq:L(z)-def}, and its dominant singularity $\rho$.

\section{A Boltzmann sampler for simply-typed terms} \label{bol}

The Boltzmann sampler approach allows us to rapidly generate random plain
$`l$-terms of sizes of order 500,000. Unfortunately, given the asymptotic
sparsity of closed simply-typed $`l$-terms in the set of plain
ones~\cite{maciej16}, a naive \emph{generate-test-reject} sampling scheme
becomes inevitably infeasible for sufficiently large term sizes.  However
remarkably, this size threshold can be postponed by interleaving the sampling
process with an optimised \emph{anticipated rejection} phase, see~\cite{BGR15},
where undesired terms are discarded as soon as it is possible to determine that
the (partially) constructed term cannot be closed nor typeable. At this point,
the whole process is interrupted and restarted.  Although in effect we obtain a
uniform sampler for closed simply-typed $`l$-terms, the power of Boltzmann
samplers is significantly constrained -- due to the fact that closed simply-typed
$`l$-terms are asymptotically negligible in the set of plain $`l$-terms, the
number of expected retrials tends to infinity as the target term size
increases.

Following our empirical experiments, we calibrated the branching probabilities
so the expected outcome size to 120 -- the currently biggest practical size
achievable. In order to find the suitable $x$, we solve numerically
Equation~\eqref{eq:E(N)} for $x$ with $\mathbb{E}_x(N) = 120$. The numerical
approximation of $x$ is then \[x \approx 0.29558095907. \]

Following the construction of Boltzmann samplers in the case of plain $`l$-terms, we have to compute three branching probabilities deciding whether the sampler generates a random de~Bruijn index, an abstraction or an application. And so
\begin{itemize}
\item the probability of constructing a de Bruijn index becomes $0.35700035696434995$,
\item the probability of a lambda abstraction becomes $0.29558095907$, and finally
\item the probability of an application becomes $0.34741868396$.
\end{itemize}
Furthermore, whenever we decide to create a de Bruijn index, the probability of constructing \verb+0+ is equal to $0.7044190409261122$, while a successor is chosen with probability
$0.29558095907$.

\subsection{Deriving a Boltzmann sampler from an exhaustive generator}

When generating all terms of a given size, the Prolog system explores
all possibilities via backtracking. For a random generator,
deterministic steps will be used instead, guided by the probabilities determined
by the Boltzmann sampling mechanism.

Our code is parametrized by the size interval for the generated random terms as well
as the maximum number of steps until the {\em being closed} and {\em being simply-typed}
constraints are both met. Moreover, the code relies on precomputed branching probabilities.
%Their values are obtained according to the recursive combinatorial
%specification of $`l$-terms by determining the appropriate complex function and evaluating it
%in the vicinity of its dominant singularity. The detailed process of computing the desired
%values is described in~\cite{grygiel2015}. In our case, it turns out that in order to construct
%a plain term of expected size $120$ the probabilities in question are as
%follows:
%\begin{itemize}
%\item the probability of constructing a de Bruijn index is $0.35700035696434995$
%\item the probability of a lambda abstraction is $0.29558095907$
%\item the probability of an application
%is $0.34741868396$.
%\end{itemize}
%Furthermore, whenever we decide to create a de Bruijn index the probability of constructing
%zero is equal to $0.7044190409261122$, while a successor is chosen with probability
%$0.29558095907$. Hence, at each step of the construction process we draw uniformly at random
%a real from the interval $[0,1]$ and on its basis we decide which constructor to add.
At each step of the construction process we draw uniformly at random
a real from the interval $[0,1]$ and on its basis, we decide which constructor to add.
\begin{code}
min_size(120).
max_size(150).
max_steps(10000000).
boltzmann_index(R):-R<0.35700035696434995.
boltzmann_lambda(R):-R<0.6525813160382378.
boltzmann_leaf(R):-R<0.7044190409261122.
\end{code}
The very high value of retries, {\tt max\_steps}, is coming from
the discussed sparsity of simply-typed terms among all plain terms.
The Boltzmann sampler can be fine-tuned via {\tt min\_size}
and {\tt max\_size} to search for terms in an interval
for which the probabilities of the sampler have been calibrated.

The predicate {\tt ranTypable} returns a term {\tt X}, its type {\tt T}
as well as the size of the term and the number of trial steps it took
to find the term. Note that the {\bf !} ensures that finding the first simply
typed term of the required size stops the search.\\

\noindent {\em {\tt ranTypable(X,T,Size,Steps)} is true if {\tt X} is is a (uniformly) randomly generated closed simply-typed lambda term of type {\tt T} and natural size {\tt Size} where {\tt Size} is a natural number and {\tt Steps} counts the number of trials needed to obtain {\tt X}.}\\

\begin{code}
ranTypable(X,T,Size,Steps):-
  max_size(Max),
  min_size(Min),
  max_steps(MaxSteps),
  between(1,MaxSteps,Steps),
    random(R),
    ranTypable(Max,R,X,T,[],0,Size0),
  Size0>=Min,
  !,
  Size is Size0+1.
\end{code}
Note that it calls the predicate {\tt random/1}, returning
a random value between 0 and 1, with the convention that
each predicate provides such a value for the next one(s) it calls,
a convention that will be consistently followed in the code.

The predicate {\tt ranTypable/7} follows the outline of the corresponding
non-deterministic generator, except that it is driven by deterministic choices
provided by the Boltzmann branching probabilities that
decide which branch is taken.

Note that the parameter {\tt Max} preempts growing a term above the
specified size interval as early as that happens.
Like in the generator, on which it is based,
type inference is interleaved with
term building. As a result, we prevent building terms with
subterms that are not simply-typed, as soon as
such a subterm is found. Note also that {\tt ``!''} is
used in each clause as a convenient way to commit to the appropriate
choice in case of success of the Boltzmann sampler.
\begin{code}
ranTypable(Max,R,X,V,Vs,N1,N2):-boltzmann_index(R),!,
  random(NewR),
  pickIndex(Max,NewR,X,Vs,V,N1,N2).
ranTypable(Max,R,l(A),(X->Xs),Vs,N1,N3):-boltzmann_lambda(R),!,
  next(Max,NewR,N1,N2),
  ranTypable(Max,NewR,A,Xs,[X|Vs],N2,N3).
ranTypable(Max,_R,a(A,B),Xs,Vs,N1,N5):-
  next(Max,R1,N1,N2),
  ranTypable(Max,R1,A,(X->Xs),Vs,N2,N3),
  next(Max,R2,N3,N4),
  ranTypable(Max,R2,B,X,Vs,N4,N5).
\end{code}

Besides ensuring that types assigned to a leaf are consistent
with the type acquired so far by their binder, the predicate
{\tt pickIndex/7} also enforces the property of being a closed
term by picking variables from the list of possible binders
above it, on the path to the root.
\begin{code}
pickIndex(_,R,0,[V|_],V0,N,N):-boltzmann_leaf(R),!,
  unify_with_occurs_check(V0,V).
pickIndex(Max,_,s(X),[_|Vs],V,N1,N3):-
  next(Max,NewR,N1,N2),
  pickIndex(Max,NewR,X,Vs,V,N2,N3).
\end{code}
Finally, the helper predicate {\tt next/4} ensures
that the size count accumulated so far is not above
the required interval, while providing a random value
to be used by the next call.
\begin{code}
next(Max,R,N1,N2):-N1<Max,N2 is N1+1,random(R).
\end{code}

\BX
A uniformly random simply-typed $`l$-term of size {\tt 137} and its type, obtained
after {\tt 1070126} trial steps in {\tt 4.388} seconds.
\begin{codex}
l(a(l(l(l(l(l(a(s(s(0)),a(l(a(l(l(l(0))),l(a(0,a(0,a(s(s(0)),
  a(l(a(l(0),a(a(l(l(l(l(s(s(s(0))))))),s(s(0))),a(0,a(0,a(l(l(0)),
  l(a(l(l(l(s(s(s(0)))))),s(0))))))))),l(0)))))))),a(0,a(s(s(0)),
  a(a(s(0),0),0)))))))))),l(a(l(a(0,a(l(l(s(0))),l(l(l(0)))))),
  l(a(l(a(0,a(l(a(l(l(l(l(s(0))))),l(s(s(0))))),l(s(0))))),a(l(l(a(l(0),
  l(a(l(l(l(a(0,a(0,l(l(0))))))),l(s(0))))))),s(s(0)))))))))

(A->B->((C->D->D)->E->F->G)->(((E->F->G)->G)->
  ((E->F->G)->G)->C->D->D)->((E->F->G)->G)->E->F->G)
\end{codex}
\EX
\section{Generating simply-typed normal forms} \label{nfgen}

Normal forms are $`l$-terms that cannot be further $\beta$-reduced. In other words, they avoid \emph{redexes} as subterms, i.e.~applications with lambda abstractions on their left branches.

\subsection{Generating normal forms of given size}
To generate normal forms we simply add to {\tt genLambda} the constraint {\tt notLambda/1}
ensuring that the left branch of an application node is anything
except an {\tt l/1} lambda node.\\

\noindent {\em {\tt genClosed(S,X)} is true if {\tt X} is is a plain lambda term in normal forma and of natural size {\tt S} where {\tt S} is a natural number in successor arithmetic.}

\begin{code}
genNF(s(S),X):-genNF(X,S,0).

genNF(X,N1,N2):-nth_elem(X,N1,N2).
genNF(l(A),s(N1),N2):-genNF(A,N1,N2).
genNF(a(A,B),s(s(N1)),N3):-notLambda(A),genNF(A,N1,N2),genNF(B,N2,N3).

notLambda(0).
notLambda(s(_)).
notLambda(a(_,_)).
\end{code}

\BX
Plain normal forms of natural size 5.
\begin{codex}
?- genNF(s(s(s(s(0)))),X).
X = s(s(s(0))) ;
X = l(s(s(0))) ;
X = l(l(s(0))) ;
X = l(l(l(0))) ;
X = l(a(0, 0)) ;
X = a(0, s(0)) ;
X = a(0, l(0)) ;
X = a(s(0), 0) .
\end{codex}
\EX
Counts for plain (untyped) normal forms, up to size 16, are given by the sequence:
\[0,1,2,4,8,17,38,89,216,539,1374,3562,9360,24871,66706,180340,490912.\]

\subsection{Interleaving generation and type inference}

Like in the case of the set of simply-typed $`l$-terms,
we can define the more efficient
combined generator and type inferrer
predicate {\tt genTypableNF/5}.\\

\noindent {\em {\tt genPlainTypableNF(S,X,T)} is true if {\tt X} is is a plain simply-typed lambda term of type {\tt T} in normal form and of natural size {\tt S} where {\tt S} is a natural number in successor arithmetic.}\\

\noindent {\em {\tt genPlainTypableNF(S,X,T)} is true if {\tt X} is is a closed simply-typed lambda term of type {\tt T} in normal form and of natural size {\tt S} where {\tt S} is a natural number in successor arithmetic.}\\

\begin{code}
genPlainTypableNF(S,X,T):-genTypableNF(S,_,X,T).

genClosedTypableNF(S,X,T):-genTypableNF(S,[],X,T).

genTypableNF(s(S),Vs,X,T):-genTypableNF(X,T,Vs,S,0).

genTypableNF(X,V,Vs,N1,N2):-genIndex(X,V,Vs,N1,N2).
genTypableNF(l(A),(X->Xs),Vs,s(N1),N2):-genTypableNF(A,Xs,[X|Vs],N1,N2).
genTypableNF(a(A,B),Xs,Vs,s(s(N1)),N3):-notLambda(A),
  genTypableNF(A,(X->Xs),Vs,N1,N2),
  genTypableNF(B,X,Vs,N2,N3).
\end{code}

\BX
Simply-typed normal forms of size 6 and their types.
\begin{codex}
?- genClosedTypableNF(s(s(s(s(s(0))))),X,T).
X = l(l(l(s(0)))),T =  (A->B->C->B) ;
X = l(l(l(l(0)))),T =  (A->B->C->D->D) ;
X = l(a(0, l(0))),T =  (((A->A)->B)->B) ;
\end{codex}
% 0,0,1,1,2,3,7,11,25,52,110,241,537,1219,2767,6439,14945
\EX

We are now able to efficiently generate counts for simply-typed normal forms of a given size.

\BX
Counts for closed simply-typed normal forms up to size 18.
\[
0,0,1,1,2,3,7,11,25,52,110,241,537,1219,2767,6439,14945,35253,83214.
\]
\EX

\section{Boltzmann sampler for simply-typed normal forms} \label{nfbol}

When restricted to normal forms, the Boltzmann sampler is derived in a similar way from
the corresponding exhaustive generator. In order to find the appropriate branching probabilities, we exploit the following combinatorial system defining the set $\mathcal{N}$ of \emph{normal forms} using the set $\mathcal{M}$ of so called \emph{neutral forms}.
\begin{eqnarray*}
\mathcal{N} &=& \mathcal{M} \mid \lambda\, \mathcal{N}\\
\mathcal{M} &=& \mathcal{M}\, \mathcal{N} \mid \mathcal{D}
\end{eqnarray*}
A normal form is either a neutral term, or an abstraction followed with a normal form. A neutral term, in turn, is either an application of a neutral term to a normal form, or a de~Bruijn index.

With this description of normal forms, we are ready to recompute the branching probabilities (see~\cite{Duchon:2004} for details) for a Boltzmann sampler generating normal forms.
Similarly as in the case of plain terms, we calibrated the branching probabilities so to set the expected outcome size to 120.

The resulting probabilities  and limits are given by the following predicates:

\begin{code}
boltzmann_nf_lambda(R):-R<0.3333158264186935. % an l/1, otherwise neutral
boltzmann_nf_index(R):-R<0.5062759837493023.  % neutral: index, not a/2
boltzmann_nf_leaf(R):-R<0.6666841735813065.   % neutral: 0, otherwise s/1
min_nf_size(60).
max_nf_size(80).
max_nf_steps(10000000).
\end{code}

The predicate {\tt ranTypableNF} generates a simply-typed term {\tt X}
in normal form and its type {\tt T}, while computing the size of the term
and the number of trial steps used to find it. \\
% Note the use of Prolog's cut ({\tt !}) operation to stop the search once the right size is reached.

\noindent {\em {\tt ranTypableNF(X,T,Size,Steps)} is true if {\tt X} is is a (uniformly) randomly generated closed simply-typed lambda term of type {\tt T} in normal form and of natural size {\tt Size} where {\tt Size} is a natural number and {\tt Steps} counts the number of trials needed to obtain {\tt X}.}
\begin{code}
ranTypableNF(X,T,Size,Steps):-
  max_nf_size(Max),
  min_nf_size(Min),
  max_nf_steps(MaxSteps),
  between(1,MaxSteps,Steps),
    random(R),
    ranTypableNF(Max,R,X,T,[],0,Size0),
  Size0>=Min,
  !,
  Size is Size0+1.
\end{code}

First, a probabilistic choice is made
between a normal form wrapped up by a lambda binder
and a {\em neutral term}.
% NF
\begin{code}
ranTypableNF(Max,R,l(A),(X->Xs),Vs,N1,N3):-
  boltzmann_nf_lambda(R),!, %lambda
  next(Max,NewR,N1,N2),
  ranTypableNF(Max,NewR,A,Xs,[X|Vs],N2,N3).
\end{code}

The choice between the next two clauses is decided
by the guard {\tt boltzmann\_nf\_index}. If satisfied,
the recursive path towards a de~Bruijn index is chosen.
Otherwise, an application is generated. Note the use of the cut
operation ({\tt !}) to commit to the first clause when its guard succeeds.
% neutral NF
\begin{code}
ranTypableNF(Max,R,X,V,Vs,N1,N2):-boltzmann_nf_index(R),!,
  random(NewR),
  pickIndexNF(Max,NewR,X,Vs,V,N1,N2). % an index
ranTypableNF(Max,_R,a(A,B),Xs,Vs,N1,N5):- % an application
  next(Max,R1,N1,N2),
  ranTypableNF(Max,R1,A,(X->Xs),Vs,N2,N3),
  next(Max,R2,N3,N4),
  ranTypableNF(Max,R2,B,X,Vs,N4,N5).
\end{code}

Finally, the choice is made between the two alternatives deciding
how many successor steps are taken until a {\tt 0} leaf is reached.
\begin{code}
pickIndexNF(_,R,0,[V|_],V0,N,N):-boltzmann_nf_leaf(R),!, % zero
  unify_with_occurs_check(V0,V).
pickIndexNF(Max,_,s(X),[_|Vs],V,N1,N3):- % successor
  next(Max,NewR,N1,N2),
  pickIndexNF(Max,NewR,X,Vs,V,N2,N3).
\end{code}

\BX A random simply-typed term of size {\tt 63}
in normal form and its type, generated after {\tt 1312485} trial steps in less than a second.
\begin{codex}
l(l(l(l(a(a(s(s(0)),l(a(0,a(l(l(s(0))),l(l(l(l(l(a(s(0),l(l(a(s(0),
l(s(0))))))))))))))),l(a(a(l(l(a(l(s(0)),a(a(a(l(s(0)),a(l(0),0)),
l(s(s(0)))),l(l(l(0))))))),0),l(0))))))))

(A->((((B->C->D->E->((((F->G)->H)->G->H)->I)->J->I)->K)->K)->
(L->((M->N->O->O)->L)->(M->N->O->O)->L)->P)->Q->R->P)
\end{codex}
\EX
As there are fewer $`l$-terms of a given size in normal form,
one may wonder why we are not reaching comparable or larger sizes
to plain $`l$-terms, where our sampler was able to generate terms
over size {\tt 120}.
An investigation of the relative densities of simply-typed
terms in the two sets provides the explanation.

The table in fig. \ref{tab} compares the changes in density for simply-typed terms and simply-typed normal forms. The first column lists the sizes of the terms.
Column {\bf A} lists the number of closed simply-typed terms of a given size.
Column {\bf B} lists the ratio between plain terms and simply-typed terms.
Column {\bf C} lists counts for closed simply-typed normal forms.
Column {\bf D} lists the ratio between terms in normal form and closed simply-typed terms in normal form.
Finally, column {\bf E} computes the ratio of the two densities
given in columns {\bf B} and {\bf D}.

\begin{figure}[H]
\begin{center}
\begin{tabular}{|r||r|r|r|r|r||}
\cline{1-6}
\cline{1-6}
  \multicolumn{1}{|c||}{{Size}} &
    \multicolumn{1}{c|}{{ {\small A: typed}}} &
  \multicolumn{1}{c|}{{ {\small B: plain/typed}}} &
  \multicolumn{1}{c|}{{ {\small C: TNF}}} &
  \multicolumn{1}{c|}{{ {\small  D: NF/TNF}}} &
  \multicolumn{1}{c||}{{ {\small E: Density ratios}}} \\
\cline{1-6}
\cline{1-6}

5  & 5 & 4.400 & 3 & 5.666 & 0.776     \\ \cline{1-6}
10 & 508 & 6.988 & 110 & 12.490 & 0.559    \\ \cline{1-6}
15 & 82,419 & 10.568 & 6,439 & 28.007 & 0.377     \\ \cline{1-6}
20 & 16,019,330 & 15.800 & 473,628 & 60.040 & 0.263  \\ \cline{1-6}

\end{tabular} \
\medskip
\caption{Comparison of the ratios of simply-typed terms and simply-typed normal forms
\label{tab}}
\end{center}
\end{figure}

The plot in fig. \ref{pnf} shows the much faster growing sparsity of
simply-typed normal forms, measured
as the ratio between plain terms and their simply-typed subset and
respectively the ratio between normal forms and their simply-typed subset,
i.e.~the results shown in columns {\bf B} and {\bf D}, for sizes
up to {\tt 20}.

Finally, the plot in fig. \ref{nfdelta} shows the ratio between these two
quantities, i.e.~those listed in column {\bf E}, for sizes up to {\tt 20}.
In both charts the horizontal axis stands for the size, while the vertical
one for the number of terms.

\FIG{pnf}{Sparsity of simply-typed terms (lower curve) vs. simply-typed normal forms (upper curve)}{0.30}{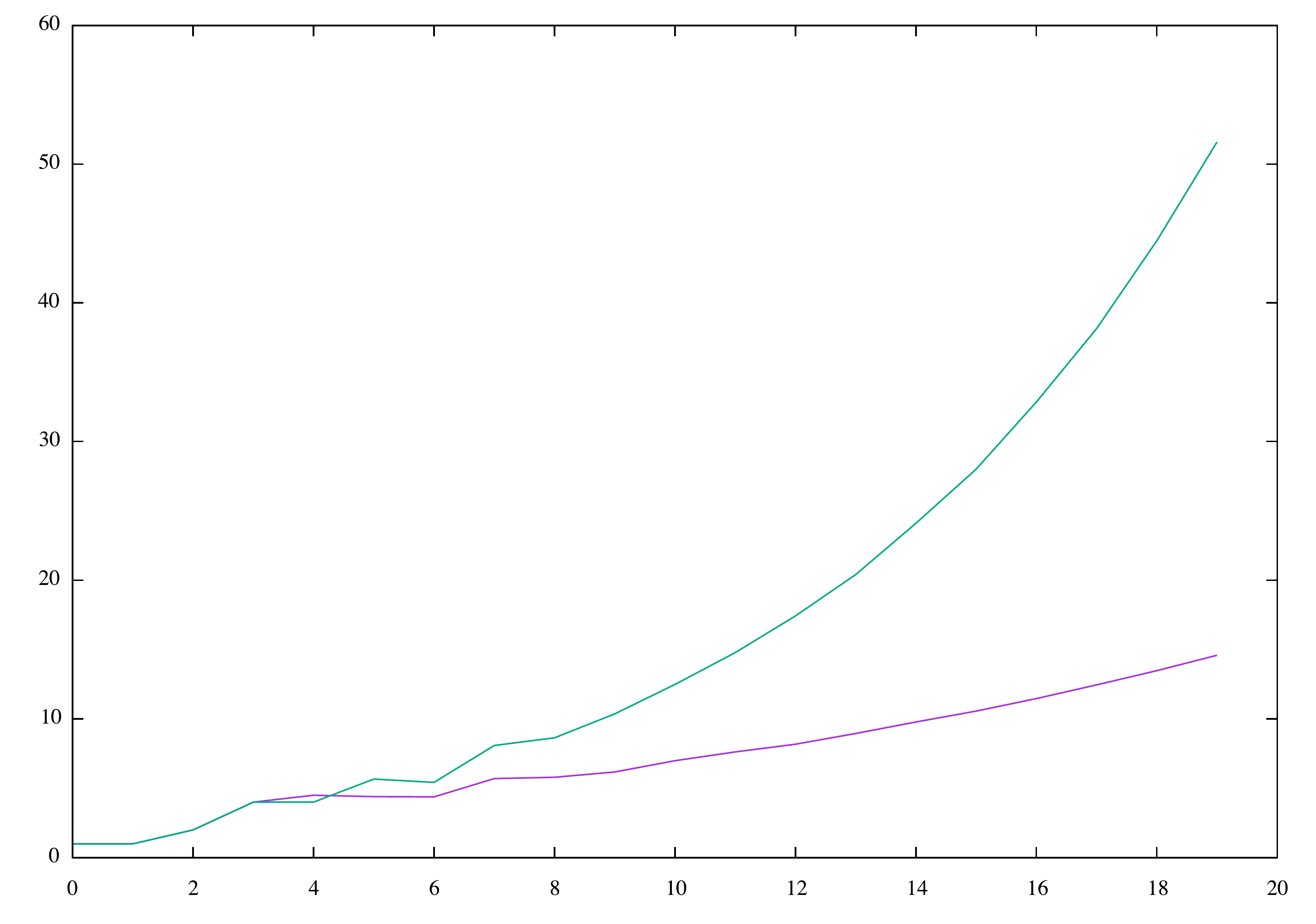}

\FIG{nfdelta}{Ratio between the density of simply-typed closed normal forms and that of simply-typed closed $`l$-terms}{0.30}{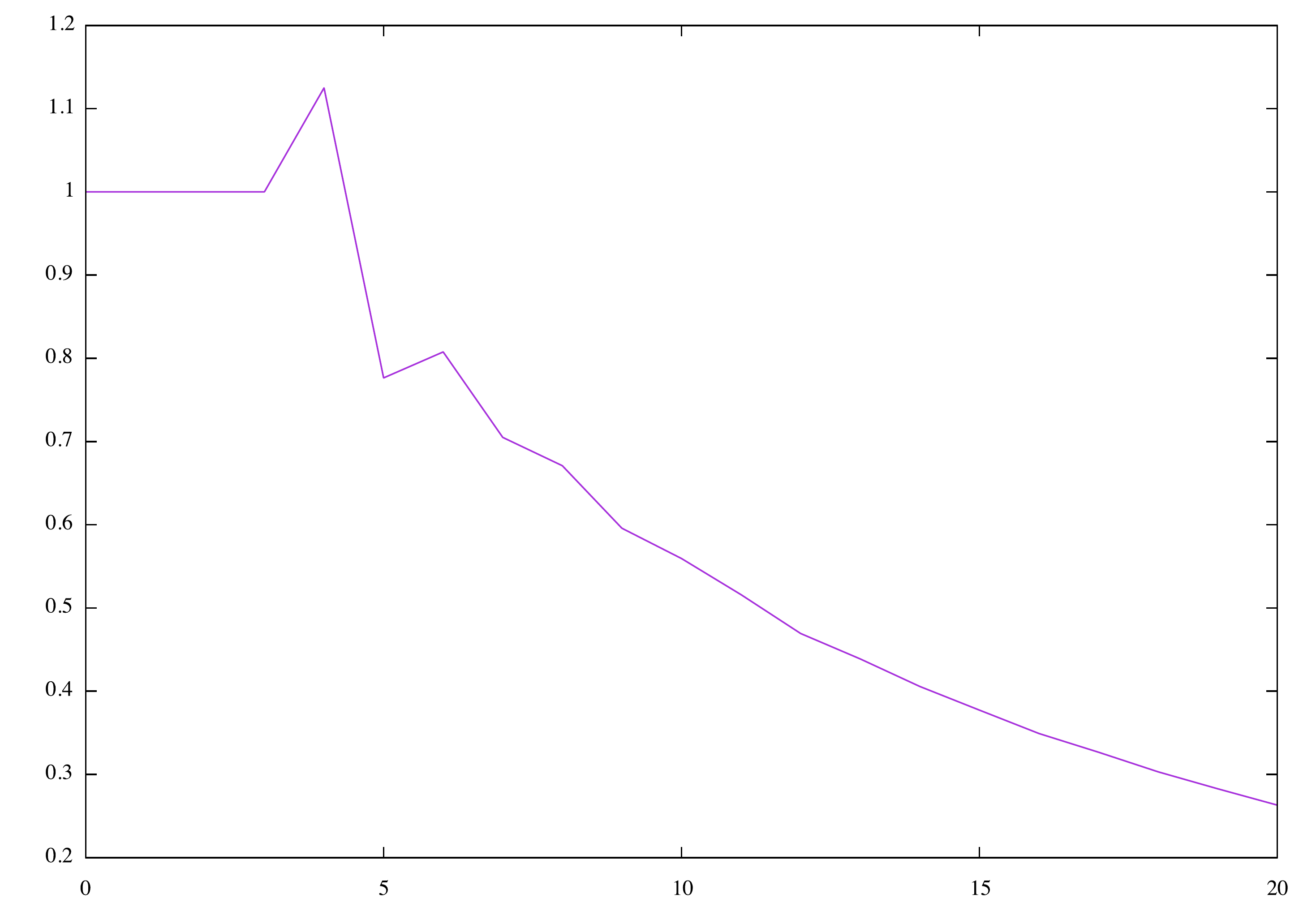}

Therefore, we see that closed simply-typed normal forms
are becoming very sparse much earlier than their plain counterparts.
While, e.g. for size $20$ there are around $1/16$ closed simply-typed
terms for each term, at the same size, for each term in normal form
there are around $1/60$ simply-typed closed terms in normal form.
As at sizes above $50$ the total number of terms is intractably high,
the increased sparsity of the simply-typed terms in normal form
becomes the critical element limiting the chances of successful search.

We leave as an open problem the study of the asymptotic behaviour of the
ratio between the density of simply-typed
closed normal forms in the set of all normal forms and the density
of simply-typed closed $`l$-terms in the set of $`l$-terms.
While our empirical data hints at the possibility that
it is asymptotically 0 for $n\rightarrow \infty$, it is still possible
to converge to a small finite limit.
Also, this behaviour could be dependent on the size definition we are using.

\begin{codeh}
% genClosedTypable/genClosedTypableNF

nf([1,1,2,4,4,5.666666666666667,5.428571428571429,8.090909090909092,8.64,10.365384615384615,12.49090909090909,14.780082987551868,17.43016759776536,20.40278917145201,24.10769786772678,28.007454573691568,32.84790899966544,38.13516580149207,44.47758790588122,51.5955629249215,60.040419907606]).

plain([1,1,2,4,4.5,4.4,4.384615384615385,5.703703703703703,5.797297297297297,6.1767676767676765,6.988188976377953,7.62582056892779,8.180624835914939,8.953135439534218,9.781185602417915,10.568048629563572,11.469698136454644,12.457100784898252,13.484006626736333,14.592226055606623,15.800088830182036]).

relrats(Rs):-plain(Ps),nf(Ns),ratios2(Ns,Ps,Rs).
\end{codeh}

 \section{Parallelizing the search} \label{par}
As multiple independent fresh restarts are used in the search for a simply-typed term or normal form,
it makes sense to run them in parallel. For multiple threads to work as efficiently as possibly on this
task, their number needs to be close to the number of actual processing elements
(cores and/or `hyper-threads', depending on the actual computer running the program).
Then, each thread can
run exactly the same code until a simply-typed term of the desired size is found, at which point all
the other threads should be terminated and the answer returned.

Implementing this model is unusually simple in SWI-Prolog, by using the predicate {\tt first\_solution},
that does exactly what we just described, for a list of goals that match the detected number
of processing elements.
The predicate {\tt multi\_gen/1} returns a simply-typed closed {\tt X}, its type {\tt T}
as well as its size and the number of steps it took to find it.
First it generates its list of goals by replicating the {\tt ranTypable/4} query
based on the number of available {\tt cpu\_count} flag.
Then, the predicate {\tt first\_solution} is called to spawn
as many threads as the number of goals. The {\tt on\_fail(continue)} property hint
ensures that one thread failing will allow the others to keep searching.
\begin{code}
multi_gen(Res):-
  Res=[X,T,Size,Steps],
  prolog_flag(cpu_count,MaxThreads),
  G=ranTypable(X,T,Size,Steps),
  length(Goals,MaxThreads),
  maplist(=(G),Goals),
  first_solution(Res,Goals,[on_fail(continue)]).
\end{code}
Our experiments on a 4 cores 8-hyper-threads Ubuntu Linux Machine, with an Intel i7 processor
have consistently returned simply-typed terms of size {\bf 140} and larger in less than a minute.
On a 44-core / 88 hyper-thread Intel Xeon machine, we have
consistently generate terms of size {\bf 180} and larger in less than a second.

Similarly, the predicate {\tt multi\_gen\_nf/1} returns a simply-typed normal form {\tt X}, its type {\tt T}
as well as its size and the number of steps it took to find it.
\begin{code}
multi_gen_nf(Res):-
  Res=[X,T,Size,Steps],
  prolog_flag(cpu_count,MaxThreads),
  G=ranTypableNF(X,T,Size,Steps),
  length(Goals,MaxThreads),
  maplist(=(G),Goals),
  first_solution(Res,Goals,[on_fail(continue)]).
\end{code}
With 8 threads running (on a 4 core / 8 hyper-threads Intel i7 machine), our experiments have consistently returned
simply-typed normal forms of size 70 and larger in less than a minute.
On a 44-core / 88 hyper-thread Intel Xeon machine, we have
consistently generated simply-typed terms in normal form
of size {\bf 80} and larger in less than a second.

\begin{codeh}
% for this, you might want to lift size limits up by 10-20
mgen:-Res=[X,T,Size,Steps],
  time(multi_gen(Res)),
  ppp(term=X),ppp(type=T),ppp(size(Size)+steps(Steps)).

mgen_nf:-Res=[X,T,Size,Steps],
  time(multi_gen_nf(Res)),
  ppp(term=X),ppp(type=T),ppp(size(Size)+steps(Steps)).
\end{codeh}

\section{Discussion}\label{disc}
An interesting open problem is whether our method can be pushed significantly farther.
We have looked into deep hashing based indexing ({\tt term\_hash} in SWI Prolog)
and tabling-based dynamic programming algorithms, using de~Bruijn terms. Unfortunately
as subterms of closed terms are not necessarily closed, even if
de~Bruijn terms can be used as ground keys, their associated
types are incomplete and dependent on the context in which
they are inferred.

While it only offers a constant factor speed-up,
parallel execution, as shown in section \ref{par}, is quite
effective in generating closed simply-typed terms of size 140
and larger and simply-typed normal forms of size 70 and higher.
This mechanism is based on independent threads, running
identical programs. Experiments with more fine-grained
execution models (e.g. allowing some sharing of generated
subterms) might need more sophisticated inter-thread
communication mechanisms, with possible changes to
the underlying runtime system.

Note also that for exhaustive generation, given the small granularity of the generation
and type inference process, the most useful parallel
execution mechanism would simply split
the task of combined generation and inference process
into a number of
disjoint sets. For instance, assuming size $n$
and $k$ {\tt l/1} constructors for $k \leq n$,
one would launch a thread exploring all possible choices,
with the remaining  $n-k$ size-units to be shared by the applications {\tt a/2}
and the weights of indices {\tt s/1}.

\section{Related work}\label{rels}

The problem of counting and generating uniformly random $`l$-terms is extensively studied in the literature.
In~\cite{normalizing13} authors considered a canonical representation of closed $`l$-terms in which variables do not contribute to the overall term size. The same model was investigated in~\cite{grygielGen}, where a sampling method based on a \emph{ranking-unranking} approach was developed.
A binary variant of lambda calculus was considered in~\cite{grygiel2015}, leading to a generation method employing Boltzmann samplers. The natural size notion was introduced in~\cite{maciej16}. The presented results included quantitative investigations of certain semantic properties, such as strong normalization or typability.

Other, non-uniform generation, approaches are also studied in the context of automated software verification.
Prominent examples include Quickcheck~\cite{claessen-quickcheck-lightweight-tool-2000} and {\scshape GAST}~\cite{Koopman2003} -- two frameworks offering facilities for random (yet not necessarily uniform) and exhaustive test generation, used in the verification of user-defined function properties and invariants.
In~\cite{palka11}  a `type-directed' mechanism for
generation of random terms was introduced,
resulting in more realistic (from the particular use case point of view) terms,
employed successfully in discovering optimization bugs in the Glasgow Haskell Compiler (GHC).
Function synthesis, given a finite set of input-output examples,
was considered in~\cite{Koopman06systematicsynthesis}. In this approach, the set of candidate functions is restricted to a subset of primitive recursive functions with abstract syntax trees defined by some context-free grammar, yielding an effective method of finding `natural' functions matching the given example set.
A statistical exploration of the structure of the
simple types of $`l$-terms of a given size
in~\cite{iclp15} gives indications that some types
frequent in human-written programs are
among the most frequently inferred ones for terms of a given size.

\section{Conclusion} \label{concl}

We have derived from logic programs
for exhaustive generation of $`l$-terms programs that
generated uniformly distributed simply-typed $`l$-terms
 via Boltzmann samplers.
 This has put at test a simple but effective
program transformation technique naturally
available in logic programming languages:
interleaving generators and constraints by
integrating them in the same predicate.
For the exhaustive generation,
we have also managed
to work within
the minimalist framework of Horn clauses with sound
unification, showing that non-trivial combinatorial problems
can be handled without any of Prolog's impure features,
except for (avoidable) uses of cuts and the use of a random generator.

Conducted empirical study of Boltzmann samplers has revealed
an intriguing discrepancy between the case of
simply-typed terms and simply-typed normal forms.
While these two classes of terms are both known to
asymptotically vanish, the significantly
faster sparsity growth of the latter
has limited our Boltzmann sampler to
sizes of order 70.

Our techniques,  combining unification of logic variables with Prolog's
backtracking mechanism, recommend logic programming as a convenient
metalanguage for the manipulation of various families of $`l$-terms and the
study of their combinatorial and computational properties. Random generation
of uniformly random simply-typed closed $`l$-terms of sizes above 120 opens the
door for potential applications in various domains of software quality assurance.  With
our generators it becomes possible to construct samplers for anonymous
functions in functional programming languages, such as Haskell or
OCaml. Since $`l$-terms in the de~Bruijn notation form in essence a nameless
template for anonymous higher-order functions, it becomes possible to introduce
alternative naming strategies and detect potential name clashes or
resolution bugs. Large closed simply-typed $`l$-terms
constitute a novel source of random functional programs and, as such, can be
useful in detecting performance or memory management issues related to data
scalability in multiple practical applications ranging from functional compilers to
automated code optimisers for $\beta$-reduction, lambda lifting or type
inference. Finally, let us note that as closed simply-typed $`l$-terms
form proofs in framework of minimal logic with a single implication type
constructor, our techniques can be applied for testing intuitionistic tautology
solvers in various proof assistants, such as Coq or Isabelle.

\section*{Acknowledgements}
We would like to thank the anonymous referees for their insightful comments and suggestions on the paper.

\bibliographystyle{acmtrans}
\bibliography{theory,tarau,proglang}

\begin{codeh}
cp(N):-counts(N,genLambda(_,_)). % 0,1,2,4,9,22,57,154,429,1223,3550
bp(N):-times(N,genLambda(_,_)).
sp(N):-show(N,genLambda(_,_)).

cc(N):-counts(N,genClosed(_,_)). % 0,0,1,1,3,6,17,41,116,313,895
bc(N):-times(N,genClosed(_,_)).
sc(N):-show(N,genClosed(_,_)).

cpt(N):-counts(N,genPlainTypable(_,_,_)). % 0,1,2,3,8,17,42,106,287,747,2069
bpt(N):-times(N,genPlainTypable(_,_,_)).
spt(N):-show(N,genPlainTypable(_,_,_)).

ct(N):-counts(N,genClosedTypable(_,_,_)). % 0,0,1,1,2,5,13,27,74,198,508
bt(N):-times(N,genClosedTypable(_,_,_)).
st(N):-show(N,genClosedTypable(_,_,_)).

cn(N):-counts(N,genNF(_,_)). % 0,1,2,4,8,17,38,89,216,539,1374
bn(N):-times(N,genNF(_,_)).
sn(N):-show(N,genNF(_,_)).

ctn(N):-counts(N,genClosedTypableNF(_,_,_)). % 0,0,1,1,2,3,7,11,25,52,110
btn(N):-times(N,genClosedTypableNF(_,_,_)).
stn(N):-show(N,genClosedTypableNF(_,_,_)).

% rpnf(N):-relcounts(N,genClosedTypable(_,_,_),genClosedTypableNF(_,_,_)).

rp(N):-relcounts(N,genLambda(_,_),genClosedTypable(_,_,_)).

rn(N):-relcounts(N,genNF(_,_),genClosedTypableNF(_,_,_)).

rgen:-time(ranTypable(X,T,Size,Steps)),
   ppp(term=X),ppp(type=T),ppp(size(Size)+steps(Steps)),
   fail.

rgen_nf:-time(ranTypableNF(X,T,Size,Steps)),
   ppp(term=X),ppp(type=T),ppp(size(Size)+steps(Steps)),
   fail.

/*
?- rgen.
% 133,503,909 inferences, 18.883 CPU in 18.915 seconds (100% CPU, 7069881 Lips)
term=l(l(l(a(l(l(a(a(l(a(l(l(l(0))),a(0,l(a(a(l(l(l(l(0)))),0),s(s(0))))))),0),a(a(s(s(s(0))),0),a(l(l(s(0))),l(a(l(a(l(0),l(l(l(l(l(a(l(a(0,l(0))),a(l(a(l(l(l(a(0,s(s(0)))))),0)),l(l(l(l(0))))))))))))),a(a(0,l(a(a(l(l(0)),l(0)),l(0)))),l(0))))))))),l(l(a(l(a(l(l(l(a(0,l(0))))),a(a(a(a(l(l(l(a(0,l(0))))),l(0)),s(0)),l(0)),a(0,a(s(s(s(s(0)))),a(l(l(0)),a(0,l(0)))))))),0)))))))
type=(A->(((B->C->D->D)->B->C->D->D)->(E->((F->G->G)->(H->H)->I)->J->K->L->M->N->((O->P->Q->R->R)->S)->S)->E->((F->G->G)->(H->H)->I)->J->K->L->M->N->((O->P->Q->R->R)->S)->S)->T->((B->C->D->D)->B->C->D->D)->U->U)
size(136)+steps(4803158)
*/
\end{codeh}

\begin{codeh}
% content of helper file stats.pl included to make this self contained

% helpers

% to unary base
n2s(0,0).
n2s(N,s(X)):-N>0,N1 is N-1,n2s(N1,X).

% constructor counters

inits(Cs):-maplist(init,Cs).

init(C):-flag(C,_,0).
inc(C):-flag(C,X,X+1).
inc(C,K):-flag(C,X,X+K).
total(C,R):-flag(C,R,R).

% counts solutions up to M
ncounts(M,Goal):-counts((=),M,Goal,_).

counts(M,Goal):-counts(n2s,M,Goal,_).

counts(Transformer,M,Goal,Ts):-
  functor(Goal,F,_),writeln(F:M),
  findall(T,(
     between(0,M,N),
       call(Transformer,N,S),
       arg(1,Goal,S),
       sols(Goal,T),
       writeln(N:T)
     ),
  Ts),
  writeln(counts=Ts),
  ratios(Ts,Rs),
  writeln(ratios=Rs).

nrelcounts(M,G1,G2):-relcounts((=),M,G1,G2).

relcounts(M,G1,G2):-relcounts(n2s,M,G1,G2).

relcounts(F,M,G1,G2):-
  counts(F,M,G1,Cs1),
  counts(F,M,G2,Cs2),
  ratios2(Cs1,Cs2,Rs),
  functor(G1,F1,_),
  functor(G2,F2,_),
  writeln(F1=Cs1),
  writeln(F2=Cs2),
  writeln(F1/F2=Rs).

% counts how many times a goal succeeds
sols(Goal, Times) :-
        Counter = counter(0),
        (   Goal,
            arg(1, Counter, N0),
            N is N0 + 1,
            nb_setarg(1, Counter, N),
            fail
        ;   arg(1, Counter, Times)
        ).

% benchmarks Goal up to M
times(M,Goal):-
  between(0,M,N),
  times1(N,Goal).

times1(N,Goal):-n2s(N,S),ntimes1(N,S,Goal).

ntimes(M,Goal):-
 between(0,M,N),
 ntimes1(N,N,Goal).

ntimes1(N,S,Goal):-arg(1,Goal,S),
  functor(Goal,F,_),
  time(sols(Goal,Count)),writeln(N:F=count:Count),
  fail.

% computes rations between consecutive terms in a sequence
ratios([X|Xs],Rs):-
  map_ratio(Xs,[X|Xs],Rs).

map_ratio([],[_],[]).
map_ratio([X|Xs],[Y|Ys],[R|Rs]):-
  (Y=:=0->R=X/Y;R is X/Y),
  map_ratio(Xs,Ys,Rs).

ratios2([],[],[]).
ratios2([X|Xs],[Y|Ys],[R|Rs]):-
  (Y=:=0->R=X/Y;R is X/Y),
  ratios2(Xs,Ys,Rs).

% generates and shows terms of size N
show(N,Goal):-
 n2s(N,S),
 nshow(S,Goal).

nshow(S,Goal):-
  functor(Goal,F,_),
  writeln(F),

  arg(1,Goal,S),
    Goal,
    show_one(Goal),
  fail.

% shows a term with renamed variables
show_one(Goal):-
  numbervars(Goal,0,_),
  Goal=..[_,_|Xs],
    member(X,Xs),
    writeln(X),
  fail
; nl.

nv(X):-numbervars(X,0,_).

ppp(X):-nv(X),writeln(X),fail.
ppp(_).

\end{codeh}

\end{document}